\documentclass[12pt]{iopart}
\usepackage{graphicx}

\begin{document}

\title{Dielectric confinement of excitons in type-I and type-II semiconductor nanorods}

\author{M. Royo, J.I. Climente, J.L. Movilla and J. Planelles }
\address{Departament de Qu\'{\i}mica F\'{\i}sica i Anal\'{\i}tica,
Universitat Jaume I, E-12080, Castell\'o, Spain}

\ead{josep.planelles@qfa.uji.es}


\begin{abstract}
We theoretically study the effect of the dielectric environment on the exciton ground 
state of CdSe and CdTe/CdSe/CdTe nanorods. We show that insulating environments enhance
the exciton recombination rate and blueshift the emission peak by tens of meV. These effects are
particularly pronounced for type-II nanorods. In these structures, the dielectric confinement may even modify the 
spatial distribution of electron and hole charges. A critical electric field is required to separate electrons
from holes, whose value increases with the insulating strength of the surroundings.
\end{abstract}

\submitto{\JPCM}

\maketitle

\section{Introduction}

Semiconductor nanocrystals are high-performance light emitters under intense
investigation because of their applications in a wide range of fields, including 
lasing technology, quantum optics, solar energy capture and biomedicine~\cite{Rogach_book}.
Due to their nanoscopic size, the electronic structure of the carriers bound in
these crystals is mainly determined by quantum confinement~\cite{ElSayedACR,LiNL}.
For this reason, recent progress in size~\cite{DabbousiJPCB}, shape~\cite{MannaJACS} and 
composition~\cite{WangNAT} control of nanocrystals has boosted their technological 
prospects~\cite{QDmarket}. Nanorods (NR) or quantum rods are a clear example of this
progress. Their elongated shape results in an anisotropic spatial confinement of
carriers which is responsible for a series of improved optical properties relative to
spherical quantum dots. These range from higher photoluminescence quantum efficiency
~\cite{PengNAT} and faster carrier relaxation~\cite{MohamedNL} to strongly polarized 
emission~\cite{HuSCI}. Furthermore, recent advances in vapor-liquid-solid methods 
have enabled the synthesis of layered semiconductor NRs~\cite{MilirionNAT,ShiehPCB,
KumarSM,SaundersChPCh}. In these systems the heterogeneous composition allows the formation of  
band structures where electrons and holes are preferably located in different spatial
regions, forming what is known as type-II quantum dots. Upon excitation, 
these systems develop a long-lived charge-separated state which makes them attractive
for photovoltaic applications~\cite{KumarSM,KlimovNAT}.

Spatial confinement is not however the only source of quantum confinement in these
structures. Nanocrystals are usually embedded in insulating materials, whose low
dielectric constant adds a severe dielectric confinement. In spherical quantum dots,
the strong isotropic confinement gives rise to similar electron and hole charge distributions.
As a result, the influence of dielectric confinement for excitons is weakened~\cite{LannooPRL,BolcattoPRB},
the main effect being an increase of the binding energy~\cite{BrusJPC,FonoberovPRB}.
One may wonder if this is also the case in NRs, where the presence of a weak 
confinement direction could lead to a different behavior.
Indeed, several studies on quasi-one-dimensional nanostructures have suggested 
the dielectric mismatch between semiconductor materials and the environment as 
the driving mechanism to explain some experimental observations. For example, we can
mention the large variation
of the optical gap of CdSe NRs compared to the transport one~\cite{KatzPRL}, the effect on
the excitonic energies observed in ZnS NRs~\cite{MandalJAP} and type-II NRs~\cite{LoJCP},
or the large magnitude of the polarization anisotropy on linear~\cite{WangSCIENCE,LanAPL,WuNL}
and nonlinear~\cite{BarzdaAPL} optical phenomena.
Dielectric confinement has also been shown to affect the dynamics of the 
electron-hole separation in type-II heterostructured NRs~\cite{KasakarageAPL} as well as 
the coupling between electrons and longitudinal optical phonons in CdSe NRs~\cite{SunPRL}. 
From the theory side, a few works have investigated excitons in dielectrically
confined CdSe nanorods, but they neglected either the longitudinal 
confinement~\cite{ShabaevNL} or the self-interaction with the polarization charges~\cite{ClimentePRB}.

In this work, we perform a theoretical study of the effects of the dielectric confinement on
the excitonic properties of semiconductor NRs. We consider homogeneous CdSe NRs 
as well as recently synthesized linear CdTe/CdSe/CdTe heterostructured NRs subject to different
dielectric environments. We use a fully 3D effective-mass and envelope-function Hamiltonian
which allows us to model sophisticated geometries. The contributions coming from the dielectric 
mismatch are accounted for using a numerical procedure, and the electron-hole correlations 
--which are important for long NRs-- are treated by carrying full configuration 
interaction (FCI) calculations. 

Our results show that in semiconductor NRs the dielectric confinement modifies the 
energy and intensity of the exciton photoluminescence. The influence is particularly
important in type-II NRs, where the asymmetry between the electron and hole charge 
distribution enables strong dielectric mismatch effects. In this kind of structures,
the electronic density shows a striking response to changes in the dielectric 
constant of the environment. In insulating environments, the enhanced electron-hole
attraction moves the electron density from the center of the NR to the CdTe/CdSe 
interfaces. 
Last, we study the effect of longitudinal electric fields on the excitonic 
states of the NRs. Our results show that a threshold field
is required to separate electrons from holes. The value of this critical field is strongly
dependent on the dielectric constant of the environment.

\section{Theory and computational details}

In the effective mass approximation the exciton Hamiltonian can be 
expressed as

\begin{equation}
\label{Hx}
H=H_\rme^0(\mathbf{r}_{\rme})+H_\mathrm{h}^0(\mathbf{r}_\mathrm{h})+V_{\mathrm{eh}}(\mathbf{r}_\mathrm{e},\mathbf{r}_\mathrm{h}),
\end{equation}

\noindent where $H_{\mathrm{e,h}}^0(\mathbf{r}_{\mathrm{e,h}})$ are single-particle Hamiltonians and
$V_{\mathrm{eh}}(\mathbf{r}_\mathrm{e},\mathbf{r}_\mathrm{h})$ is the electron-hole Coulomb interaction. 
To describe the single-particle spectra we assume the following Hamiltonian in
cylindrical coordinates and atomic untits

\begin{equation}
\label{H0}
H_\mathrm{i}^0=-\frac{1}{2 m_\mathrm{i}^*}\nabla^2_\mathrm{i}+V_\mathrm{i}^\mathrm{c}(\rho_\mathrm{i},z_\mathrm{i})
+V^{\mathrm{sp}}(\rho_\mathrm{i},z_\mathrm{i})-q_\mathrm{i}Fz_\mathrm{i}. 
\end{equation}

\noindent Here $\mathrm{i=e,h}$ is a subscript denoting electron or hole respectively, $m_\mathrm{i}$ is the 
effective mass that we assume to be constant in the whole system, $V_\mathrm{i}^\mathrm{c}(\rho_\mathrm{i},z_\mathrm{i})$ is
the step-like spatial confining potential, and $V^{\mathrm{sp}}(\rho_\mathrm{i},z_\mathrm{i})$ is the self-polarization
potential arising from the interaction of each carrier with its own polarization charges, generated
on the NR interface as a consequence of the dielectric constant mismatch with the environment.
The last term of the Hamiltonian (\ref{H0}) describes the effect of an electric field $F$ applied 
along the NR longitudinal axis, with $q_\mathrm{i}$ standing for the electric charge of the carrier.

Exciton energies and wave functions are obtained by means of FCI calculations, i.e., 
as the eigenvalues and eigenfunctions of the projection of Hamiltonian (\ref{Hx}) 
onto the two-body basis set of all possible Hartree electron-hole products.
Since the low-energy single-particle spectrum of large aspect ratio NRs only includes 
orbitals with zero azimuthal angular momentum~\cite{PlanellesPRB} we use a single-particle
basis set of 1s-gaussian functions  

\begin{equation}
\label{gauss}
 g_{i,\mathrm{x}}({\mathbf r})=\mathrm{exp}\left[-\alpha_\mathrm{x} ({\mathbf r}-{\mathbf R}_i)^2\right],
\end{equation}

\noindent to obtain the exciton energies and wave functions. The exponents $\alpha_\mathrm{x}$ 
($\mathrm{x=e,h}$ for electron and hole, respectively) are fitted variationally 
in a sphere calculation where a single gaussian function is employed. 
The gaussian functions are radially centered and equally spaced along the NR longitudinal 
axis, i.e., $\mathbf{R}_i=z_i\mathbf{k}.$  We employ a large enough number of gaussian 
functions per nm of the NR axis to saturate the space and guarantee energy convergence
\footnote{A numerical basis set formed from the single-particle
Hamiltonian eigenfunctions~\cite {ClimentePRB} would be better adapted to the
spatial confinement and hence would yield lower exciton energies, closer to
experimental values.~\cite {YuCM03,liNL01} However, this energy difference is an 
approximately constant shift for all considered NR lengths and electric fields and it does 
not affect the exciton binding energies, dipole moments and wave functions 
studied here. We have chosen to use
equidistant floating gaussians because, in contrast to the numerical
eigenfunctions, they enable a uniform saturation along the NR as well as a
continuously homogeneous description of the system, from the spherical limit
to the extremely elongated one.}.
Once the set of gaussian functions are obtained, we proceed to a symmetric orthogonalization
in order to reach a set of orthonormal functions which most closely resemble the original
basis set, both for electrons and holes. Then we build up all possible Hartree electron-hole
products that expand the FCI space in which Hamiltonian (\ref{Hx}) 
is projected.

In order to calculate the electron-hole interaction term (the electron-hole exchange 
is neglected as it does not influence the reported trends) of the FCI matrix
elements 

\begin{equation}
\label{matel}
\langle \phi_i^\mathrm{e} \phi_j^\mathrm{h}|V_{\mathrm{eh}}|\phi_k^\mathrm{e} \phi_l^\mathrm{h}\rangle ,
\end{equation}

\noindent we first obtain an electron charge density $\eta(\mathbf{r}_\mathrm{e})=\phi_i^{\mathrm{e}\,*} 
\phi_k^\mathrm{e}$ and then calculate the electrostatic potential that this charge distribution
generates onto the hole. To calculate this potential in a medium with spatially inhomogeneous
dielectric constant $\varepsilon(\mathbf{r})$, we rewrite the Poisson equation in terms
of the source charges plus the induced polarization charges:

\begin{equation}
\label{eqICC}
\mathbf{\nabla}^2 V(\mathbf{r}_\mathrm{h})= - 4\pi \,
[ \eta (\mathbf{r}_\mathrm{e}) + \eta_{\mathrm{p}}(\mathbf{r}_\mathrm{e}) ].
\end{equation}

\noindent Here $\eta_\mathrm{p} (\mathbf{r}_\mathrm{e})$ is the polarization charge density, which we
calculate with a method~\cite{MovillaCPC} equivalent to the \textit{induced charge computation} one
proposed by Boda et al.~\cite{BodaPRE} The self-polarization potential appearing in
the single-particle Hamiltonian (\ref{H0}) is calculated following a similar scheme but taking a
point source charge and scaling the potential by a factor $0.5$ due to
the self-interaction nature of this term. We refer the reader to reference 32 
for further details on the inclusion of these contributions. 

In addition to energy and carrier density distribution, we calculate the ground state electron-hole
recombination probability and electric dipole moment. For the first one, we 
use the dipole approximation and Fermi golden rule~\cite{Pawel_book}

\begin{equation}
 P \propto \left| \sum_{ij} c_{ij} \, \langle \phi_i^\mathrm{e} | \phi_j^\mathrm{h} \rangle \right|^2 p_0(T).
\end{equation}
 
\noindent Here $c_{ij}$ are the exciton ground state FCI expansion coefficients, $\phi_i^\mathrm{e}$ and $\phi_j^{\mathrm{h}}$ 
are symmetrically orthogonalized gaussian functions whose Hartree products constitute the
basis set for the FCI expansion and $\langle \phi_i^\mathrm{e} | \phi_j^\mathrm{h} \rangle$ the corresponding overlap.
Since we deal with large aspect ratio NRs in which the energy separation between the 
ground state and the low lying excited states is just a few meV, to compute the exciton ground state
recombination probability we consider thermal population effects. To this end, we assume the Boltzmann
distribution  $p_l(T)=N (g_l/g_0) \mathrm{exp}\left(-\Delta E_l/kT\right)$ for the exciton states occupation  at temperature
$T$, with $g_l$ ($g_0$) as the degeneracy factor of the state $l$ (ground state), $\Delta E_l$ the
energy difference between the state $l$ and the ground state, and $k$ the Boltzmann constant.
$N$ is the normalization constant, which ensures that the sum of all exciton states population is equal to one.
Finally, for simplicity, we omit the influence of local fields induced by the dielectric mismatch on
the exciton-photon interaction. One can check that their influence in nanorods~\cite{ShabaevNL} is
qualitatively the same as that resulting from the polarization charges we investigate.

On the other hand, we calculate the electric dipole moment as

\begin{equation}
\mu=\int[\rho^\mathrm{h}-\rho^\mathrm{e}]\,z\,dv,
\end{equation}

\noindent where $\rho^{\mathrm{e,h}}$ are the electron and hole ground state densities.

\section{Results and discussion}

\subsection{Type-I NRs}

\begin{figure}[p]
\begin{center}
\includegraphics[width=7.5cm]{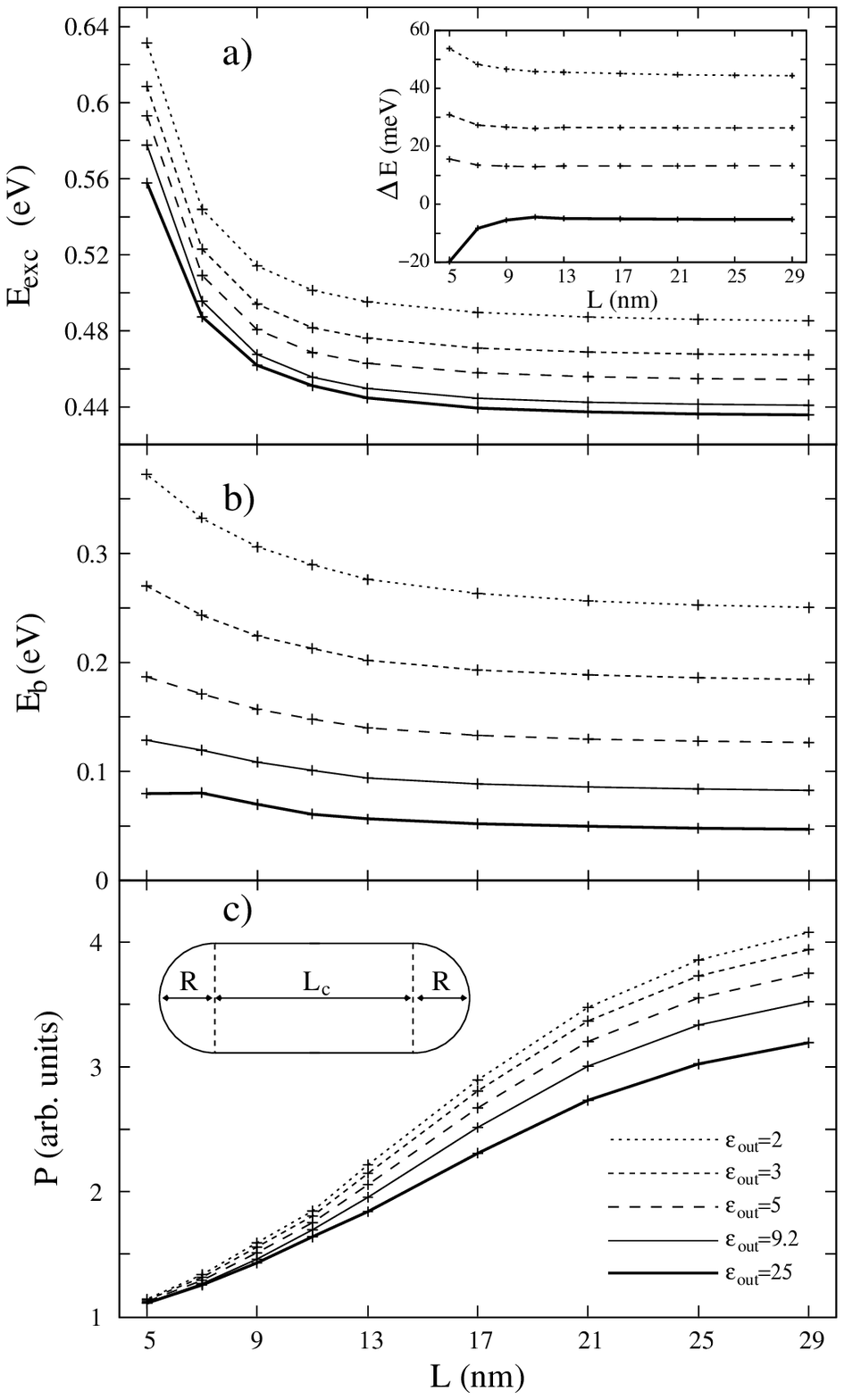}
\caption{a) Exciton ground state energies (relative to the bulk CdSe gap), b) binding energies and c) recombination
probabilities ($T=30$ K)  in homogeneous NRs with variable length $L$ 
embedded in different dielectric media. Crosses correspond to calculations. Lines are guides to the eyes. 
Different line shapes correspond to different dielectric constants. The correspondence is shown in the bottom panel.
Upper inset: Exciton energy differences between the cases with $\varepsilon_{\mathrm{out}}\neq\varepsilon_{\mathrm{in}}$ and the case 
$\varepsilon_{\mathrm{out}}=\varepsilon_{\mathrm{in}}$. Lower inset: schematic of the NR geometry.}
\label{Fig1}
\end{center}
\end{figure}

We start by studying homogeneous CdSe NRs of different lengths. The rods are composed of
a cylinder with radius $R=2$ nm and length $L_\mathrm{c}$, attached to two hemispherical
caps of radius $R=2$ nm at the extremes, yielding a total length $L=2R+L_\mathrm{c}$ (see figure~\ref{Fig1} inset). 
CdSe material parameters are used~\cite{LaheldPRB}. Thus, electron and
hole effective masses are $m_\mathrm{e}^*=0.13$ and $m_\mathrm{h}^*=0.4$. The latter corresponds to the longitudinal
mass of a light-hole, since the hole ground state in long NRs is essentially a light-hole~\cite{KatzPRL}, 
For this system, the variational gaussian coefficients are $\alpha_\mathrm{e}=0.0016$ and $\alpha_\mathrm{h}=0.0020$. 
The dielectric constant inside the NR is fixed to $\varepsilon_{\mathrm{in}}=9.2$, while outside $\varepsilon_{\mathrm{out}}$ is
varied in a wide range, in order to simulate the effect of surrounding media with different insulating strength. 
Carriers are confined inside the NR by a typical potential barrier of 4 eV.

Figure~\ref{Fig1}(a) represents the exciton ground state energy as a function of the NR length $L$ for 
embedding media of different insulating strength. For a given environment, we see that the exciton 
initially experiences a significant energy stabilization, 
 and an asymptotic value is finally attained.
This behavior, which has been observed in optical and tunneling gap measurements~\cite{KatzPRL,SteinerNL},
reflects the relaxation of the longitudinal spatial confinement. 
The asymptotic regime is usually identified with a quasi-1D system, where only radial 
confinement is present, and it explains the success of quasi-1D models in reproducing experimental observations~\cite{ShabaevNL}.

A similar relaxation is observed in figure~\ref{Fig1}(b) for the exciton binding energy as the NR is
elongated. The plot also reproduces the effect of the dielectric environment previously observed in 
spherical and cubic nanocrystals~\cite{BrusJPC,FonoberovPRB}, i.e., due to the polarization of the 
Coulomb interaction, low dielectric constant environments increase the electron-hole
attraction, and hence, the binding energy.

Despite this gain in binding energy, figure~\ref{Fig1}(a) reveals that insulating environments blueshift the exciton energy
by up to 50 meV~\cite{MovillaCPC}.
This result is driven by the self-polarization interaction and can be interpreted as follows.
Due to the dielectric mismatch, the confined carriers induce polarization charges on the NR surface.
When the NR is embedded in a medium of lower (higher)
dielectric constant, $\varepsilon_{\mathrm{in}}>\varepsilon_{\mathrm{out}}$ ($\varepsilon_{\mathrm{in}}<\varepsilon_{\mathrm{out}}$), 
the sign of the induced charges is the same (opposite) as that of the source charges.
This means that the self-interaction between source and induced charges, $V^{\mathrm{sp}}$, 
is repulsive (attractive). Conversely, the electron-hole Coulomb polarization interaction 
is attractive (repulsive). While these two contributions tend to compensate each other~\cite{BrusJPC,FonoberovPRB},
the cancelation is not exact. In all the cases we study, the self-interaction term prevails.
For insulating environments ($\varepsilon_{\mathrm{in}}>\varepsilon_{\mathrm{out}}$), this translates into
a blueshifted exciton. 

Note that the blueshift in figure~\ref{Fig1}(a) does not contradict the large reduction of the 
optical gap observed experimentally in dielectrically confined NRs~\cite{KatzPRL,ShabaevNL}.
This is because the optical gap was compared with the transport gap. Both gaps are subject to 
the self-interaction potential, but only the optical one includes electron-hole Coulomb 
polarization effects.

The inset in figure~\ref{Fig1}(a) shows the difference between the exciton energy with and
without dielectric mismatch as a function of the NR length. The energy difference
first decreases, and it becomes mostly insensitive to the length once the aspect ratio 
is larger than two. The initial decrease is due to the relaxation of the longitudinal
(dielectric) confinement, and the plateau that follows suggests that the weaker
confinement barely affects the balance between self-interaction and Coulomb polarization.

We next investigate the effect of the dielectric environment on the electron-hole recombination probability.
The results obtained at $T=30$ K are illustrated in figure~\ref{Fig1}(c). It follows from the figure that (i) the
recombination probability increases with the NR length, (ii) the dielectric confinement enhances this
probability and (iii) this enhancement is larger for long NRs.
All these results can be rationalized in terms of the strong correlation regime induced by the 
softened spatial and the dielectric confinements~\cite{ClimentePRB}. 
In all cases, for long rods thermal population of excited states becomes important and the
recombination probability saturates towards the quantum wire limit.

\subsection{Type-II NRs}

\begin{figure}[p]
\begin{center}
\includegraphics[width=7.5cm]{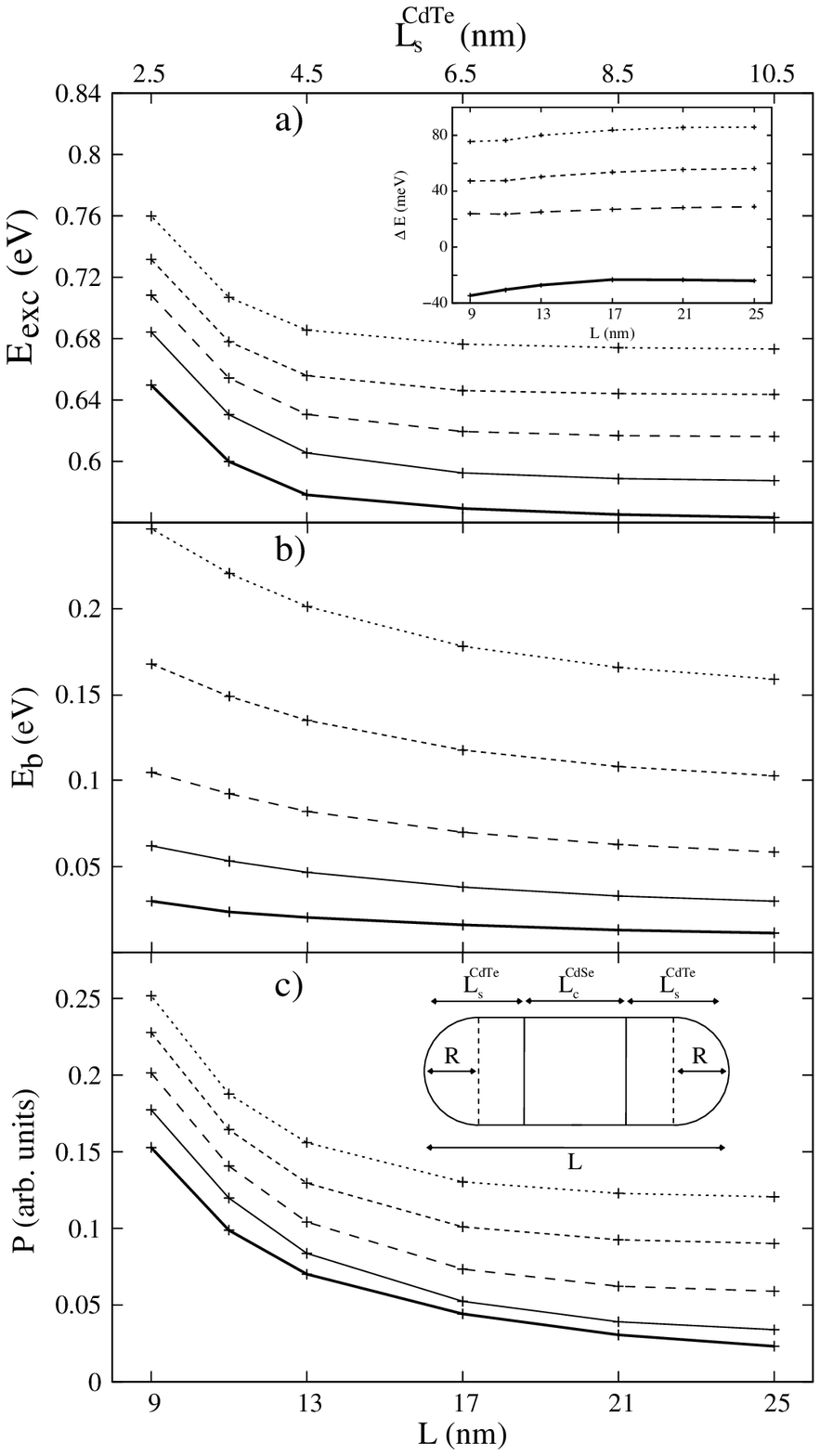}
\caption{Same as figure~\ref{Fig1} but for type-II NRs.}
\label{Fig2}
\end{center}
\end{figure}

In this section we study heterogeneous NRs similar to those synthesized in references 13 and 22.  
The rods are composed of a central CdSe cylinder (core) of radius
$R=2$ nm and length $L_\mathrm{c}^{\mathrm{CdSe}}$ attached to two external shells of CdTe. The shells 
in turn are formed by a hemispherical cap of radius $R=2$ nm and a cylinder of 
length $L_\mathrm{c}^{\mathrm{CdTe}}$ (see figure~\ref{Fig2}(c) inset). Bringing all the parts together yields 
two shells of length $L_\mathrm{s}^{\mathrm{CdTe}}=R+L_\mathrm{c}^{\mathrm{CdTe}}$ and a total NR length 
$L=2L_\mathrm{s}^{\mathrm{CdTe}}+L_\mathrm{c}^{\mathrm{CdSe}}$. These heterostructured systems are known to display a
type-II band alignment~\cite{ShiehPCB,KumarSM,SaundersChPCh,LoJCP}, where electrons are preferably located in
CdSe regions and holes in CdTe regions. To reproduce this situation, in our calculations
we include a band offset in the interface between both materials. For electrons we take
a band offset of 0.42 eV and for holes we take an inverse band offset of 0.57 eV~\cite{LeeSEMSC}.
Since the material parameters of CdSe and CdTe do not offer significant differences, we
take CdSe effective mass and dielectric constant for the whole NR. Thus, we just
consider the dielectric interface between the whole NR and the external matrix.

In figure~\ref{Fig2}(a) we show the exciton ground state energies for type-II NRs composed
by a CdSe core of length $L_\mathrm{c}^{\mathrm{CdSe}}=4$ nm and CdTe shells of increasing length
$L_\mathrm{s}^{\mathrm{CdTe}}$. Different embbeding media are considered. As in the case of homogenous
NRs, for a given environment the exciton experiences an initial energy stabilization
and later it reaches an asymptotic value. Also, the same qualitative response to the 
dielectric environment is observed. However, the magnitude of the energy shifts originated by
the dielectric confinement is now about twice that of type-I NRs, reaching values
as large as 100 meV (see figure~\ref{Fig2}(a) inset).
The reason is that the spatial separation of electron and hole charge distributions
in type-II nanostructures weakens the Coulomb polarization term, as reflected in
the smaller binding energies displayed in figure~\ref{Fig2}(b),
but not the self-polarization. This leads to greatly enhanced dielectric mismatch effects.

At this point it is worth noting that the effect of the dielectric confinement 
predicted in figure~\ref{Fig2}(a) is consistent with the main trends reported in 
reference 22, 
where the photoluminescence spectra of similar 
CdTe/CdSe/CdTe NRs were compared for solvents with different dielectric 
constant. A blueshift of the exciton emission energy by tens of meV was
observed under low dielectric constant environments (figure~7 in their work).
This confirms the prevalence of the self-interaction potential over the electron-hole
Coulomb one. The irregular differences between the energy shifts originated by the 
two low dielectric constant solvents of reference 22 are probably connected with
microscopic effects, which are beyond our continuum model.

The inset in figure~\ref{Fig2}(a) shows the difference between the exciton energy with and
without dielectric mismatch as a function of the NR length. As in type-I NRs, the 
increasing anisotropy has a weak influence.

Figure~\ref{Fig2}(c) shows the electron-hole recombination probability of type-II NRs at $T=30$ K.
As can be observed, the probability is much smaller than in type-I
NRs due to the charge separation, which was already noted in related experiments~\cite{ShiehPCB}.
In addition, contrary to type-I NRs (figure~\ref{Fig1}(c)), the recombination 
probability now decreases with the NR lenght. This is because the length increase 
comes from longer CdTe shells, so that the hole lies further away from the electron,
which leads to an additional reduction of the electron-hole overlap. 
The effect of the dielectric environment is also quite different from the homogeneous 
NR case. Insulating environments still enhance the recombination probability, but:
(i) the enhancement does not vary with $L$, because the size increase of the 
CdTe shells does not entail an increase in the role of the electron-hole correlations,
and (ii) the relative enhancement is many times larger. For example, at $L=25$ nm the 
recombination probability for $\varepsilon_{\mathrm{out}}=2$ is $\sim 3.5$ times that of 
$\varepsilon_{\mathrm{out}}=9.2$, compared to $\sim 1.2$ times in type-I NRs.
This is another manifestation of the important role of dielectric mismatch in type-II
structures.

\begin{figure}[p]
\begin{center}
\includegraphics[width=7.5cm]{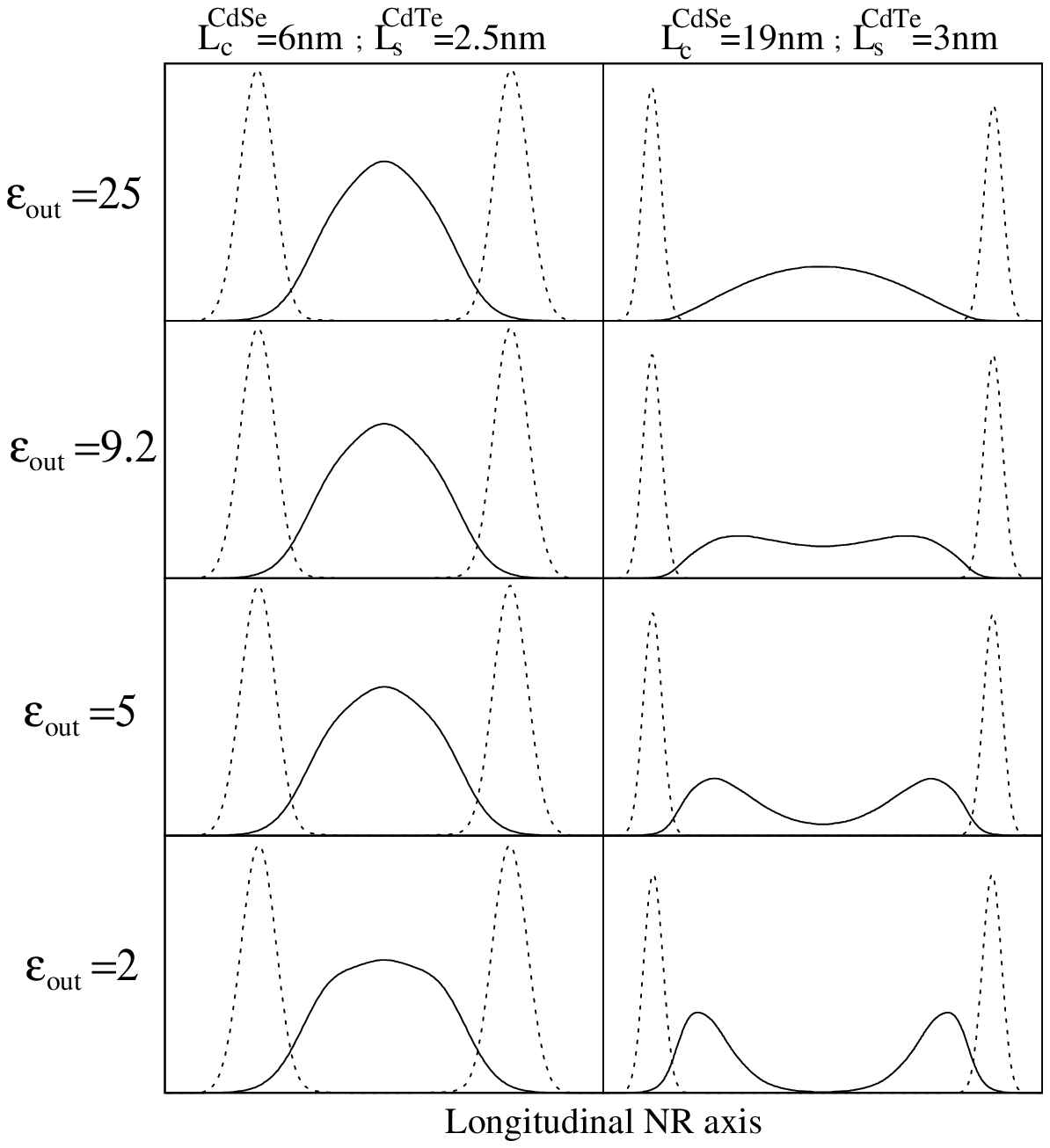}
\caption{Electron (solid lines) and hole (dashed lines) densities along the longitudinal axis, for type-II NRs of 
$L_\mathrm{c}^{\mathrm{CdSe}}=6$ nm and $L_\mathrm{s}^{\mathrm{CdTe}}=2.5$ nm (left), and
$L_\mathrm{c}^{\mathrm{CdSe}}=19$ nm and $L_\mathrm{s}^{\mathrm{CdTe}}=3$ nm (right).
The dielectric constants of the surroundings are indicated on the left of each row.}
\label{Fig3}
\end{center} 
\end{figure}

Next we show that the strong influence of dielectric confinement in type-II NRs
may even reshape the exciton wavefunction. Figure~\ref{Fig3} illustrates the electron 
(solid line) and hole (dashed line) density profiles along the NR longitudinal axis. 
Left (right) panels correspond to NRs of dimensions $L_\mathrm{c}^{\mathrm{CdSe}}=6$ nm and 
$L_\mathrm{s}^{\mathrm{CdTe}}=2.5$ nm ($L_\mathrm{c}^{\mathrm{CdSe}}=19$ nm and 
$L_\mathrm{s}^{\mathrm{CdTe}}=3$ nm) embedded in media of 
different insulating strength. No noticeable effects arise in the case of the shorter NR. 
By contrast, as the longer NR is embedded in strong insulating media, the electron 
moves from the rod center to the CdTe shells.
For a strong enough dielectric mismatch, the electron density even develops a deep 
valley at the center of the NR (see e.g. $\varepsilon_{\mathrm{out}}=2$, bottom right panel in
figure~\ref{Fig3}). The driving force of this behavior is the increase of the 
electron-hole interaction by means of the polarization charges.
As the CdSe core is elongated, this attractive potential starts dominating
over the longitudinal spatial potential felt by the electron, which is then dragged by
the hole towards the material interface. This phenomenon is favored for long
CdSe cores and short CdTe shells. 

The electron localization near the external shells evidences a regime where the 
role of the longitudinal spatial confinement is taken over by the dielectric 
confinement. Moreover, important implications follow from this phenomenon, such as
enhanced sensitivity of the exciton near the CdSe/CdTe interface and reduced coupling 
to impurities and defects in the center of the rod. 

\subsection{Electric field effect}

In the last few years both theoretical~\cite{LiPRB} and experimental~\cite{MullerNL,RothenbergNL}
studies have pointed out interesting properties for technological devices arising from the 
application of an external electric field along the longitudinal direction of NRs. 
The electric field separates electrons from holes, thus reducing the radiative 
recombination probability. The rate at which this happens is known to be affected by the 
quantum confinement, which is related to the quantum confined Stark effect.
Having observed the strong influence of dielectric confinement in NRs at zero field, 
we next probe how it modifies the exciton response to longitudinal electric fields.

\begin{figure}[p]
\begin{center}
\includegraphics[width=7.5cm]{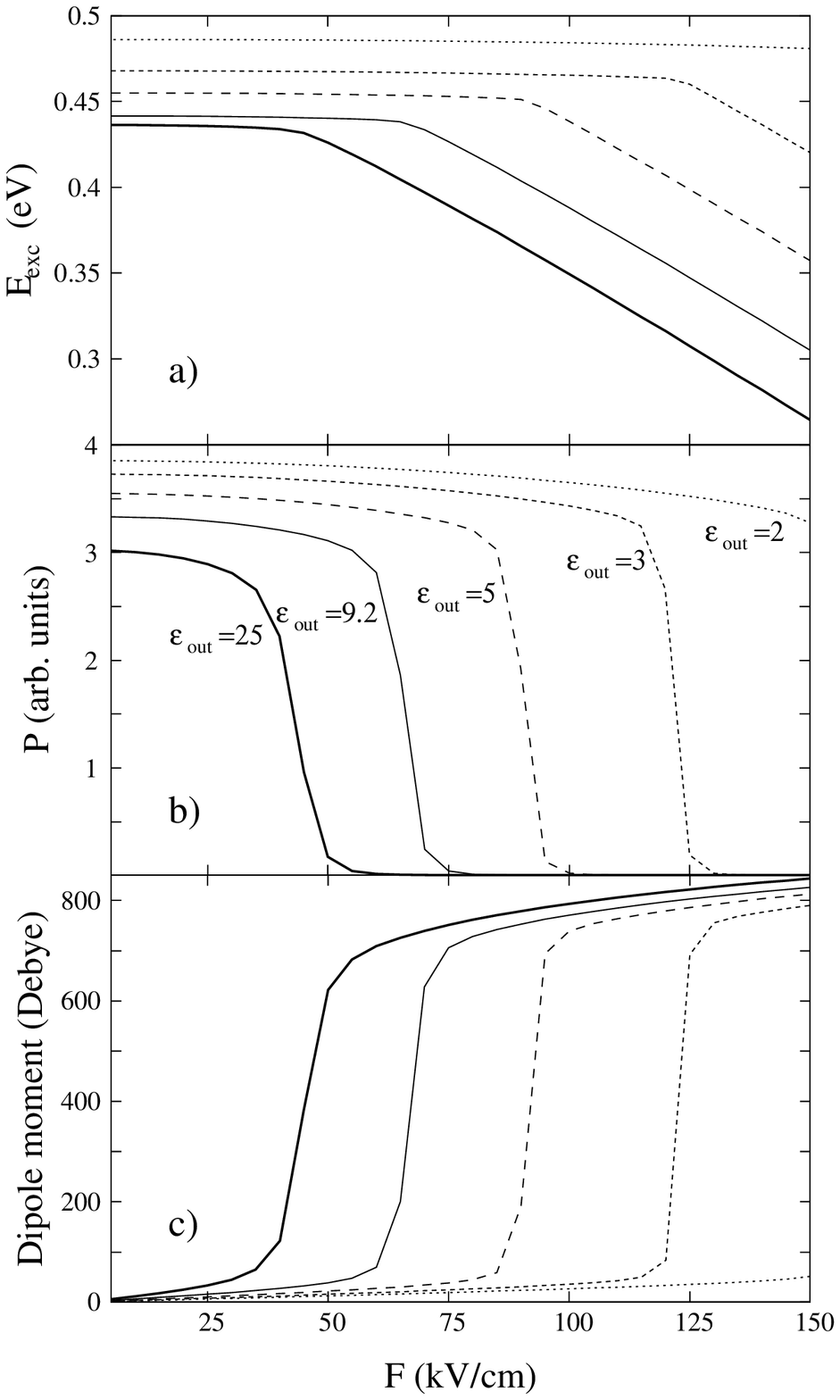}
\caption{a) Exciton ground state energies (relative to the bulk CdSe gap), 
b) recombination probabilities ($T=30$ K) and c) dipole moments of a $L=25$ nm homogeneous NR vs.~the 
applied electric field. The dielectric constants of the media are indicated by the lines in panel b).}
\label{Fig4} 
\end{center}
\end{figure}

In figure~\ref{Fig4} we study the electric field effect over the exciton ground state energy (a),
electron-hole recombination probability ($T=30$ K) (b) and dipole moment (c), for a
homogeneous CdSe NR of length $L=25$ nm in different media. 
As can be seen, there is a critical electric field from which the system evolves in a different way. 
This is the field required to induce the electron-hole separation. 
The observed strong shift is characteristic of large aspect ratio NRs, where carriers 
barely feel the spatial longitudinal confinement, so that the electric field 
almost only competes with the Coulomb attraction. Note the contrast of this 
behaviour and that of nanocrystals with strong longitudinal confinement, where 
a rather gradual response to the electric field is found (see, e.g., 
reference~\cite{MenendezPRB04} and references therein).
The electron-hole separation is reflected by a redshift of the exciton energy (figure~\ref{Fig4}(a)), a 
sudden reduction of the exciton recombination probability (figure~\ref{Fig4}(b)) 
and an abrupt increase of the dipole moment (figure~\ref{Fig4}(c)).
The abrupt response to the electric field is consistent with the rapid
switches observed in optical spectroscopy experiments~\cite{RothenbergNL}.

Figure~\ref{Fig4} proves that the dielectric confinement has important 
effects on the exciton response to electric fields. 
The critical field required to separate electrons from holes increases 
significantly with the insulating strength of the environment.
This is due to the abovementioned modulation of the exciton binding energy.

\begin{figure}[p]
\begin{center}
\includegraphics[width=7.5cm]{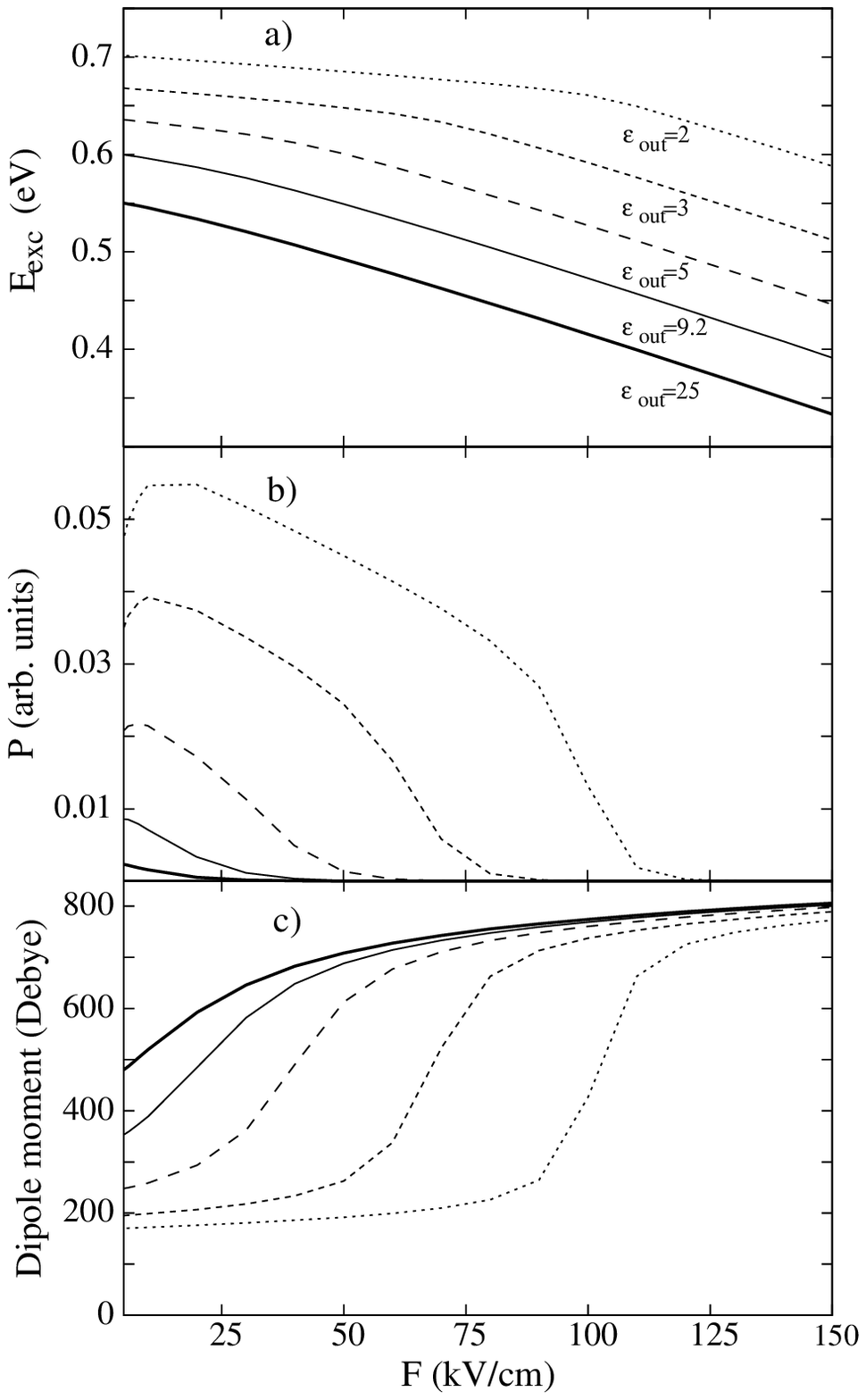}
\caption{Same as figure~\ref{Fig4} but for a type-II NR with $L_\mathrm{c}^{\mathrm{CdSe}}=19$ nm
and $L_\mathrm{s}^{\mathrm{CdTe}}=3$ nm.}
\label{Fig5}
\end{center}
\end{figure}

We next illustrate the electric field effect on type-II NRs. Results are shown in figure~\ref{Fig5} 
for a NR of $L_\mathrm{c}^{\mathrm{CdSe}}=19$ nm and $L_\mathrm{s}^{\mathrm{CdTe}}=3$ nm (total length $L=25$ nm). 
The same trends as in homogeneous NRs are observed, but now, since the
electron-hole interaction is weaker, smaller fields are required to separate both particles
and this process takes place more gradually. In any case, the influence of the dielectric 
environment on the exciton response to electric fields is still felt, and it can increase the 
critical field value over an order of magnitude.
The anomalous evolution observed at small fields in the recombination probability (figure~\ref{Fig5}(b))
is explained as follows. The electric field breaks the double-degeneracy of the hole states localized
in the CdTe caps. Since in figure~\ref{Fig5}(b) we just show the ground state recombination probability,
the initial increase comes from the thermal depopulation of the first excited state 
in favor of the ground state.
  
\begin{figure}[p]
\begin{center}
\includegraphics[width=7.5cm]{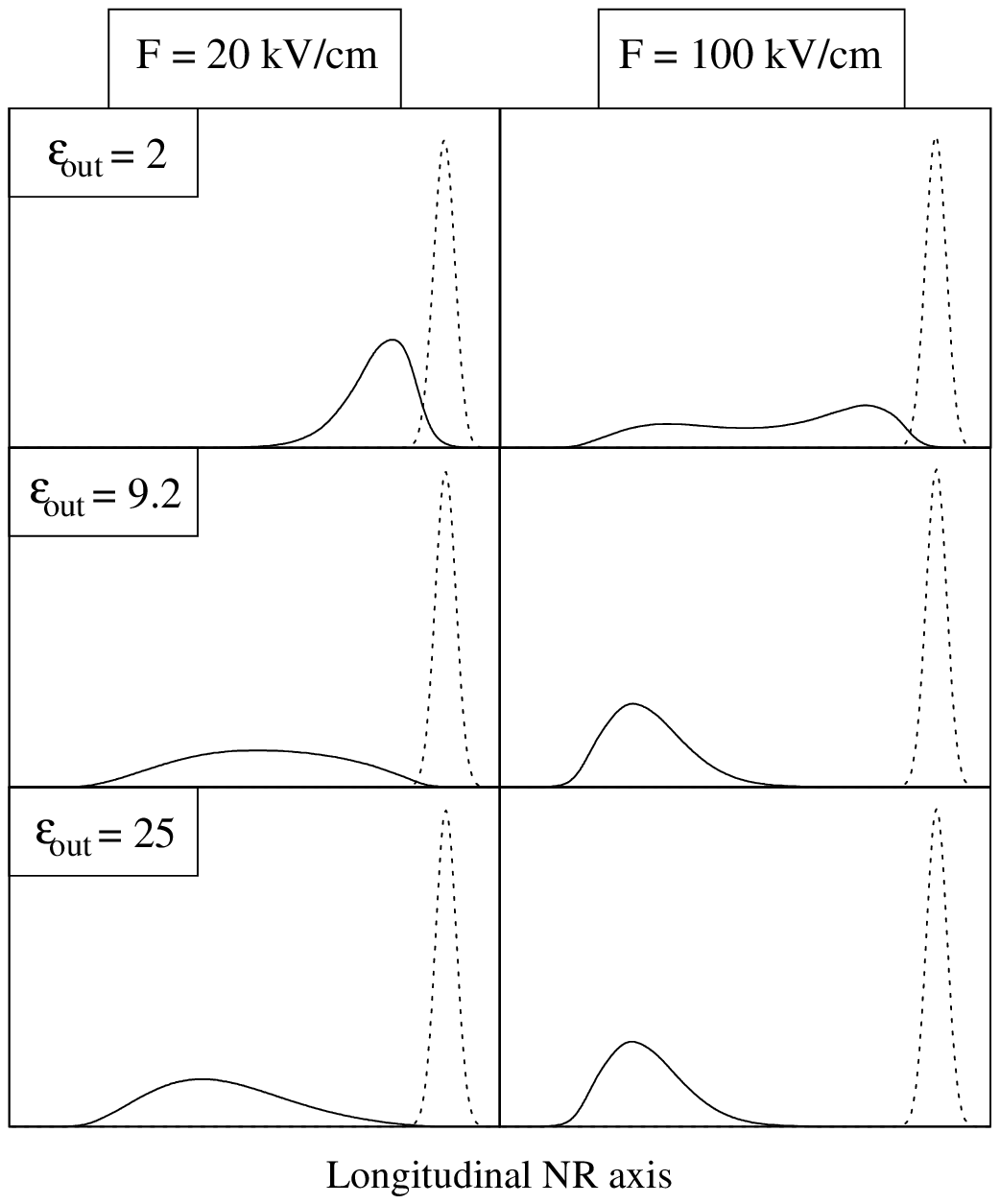}
\caption{Electron (solid lines) and hole (dashed lines) densities along the longitudinal axis, for a type-II NR 
of $L_\mathrm{c}^{\mathrm{CdSe}}=19$ nm and $L_\mathrm{s}^{\mathrm{CdTe}}=3$nm subject
to electric fields of 20 and 100 kV/cm. The dielectric constants of the different media are enclosed on the top-left corner of each row.} 
\label{Fig6}
\end{center}
\end{figure}

Finally, we focus our attention on the evolution of the exciton charge density under the
influence of electric fields. In homogeneous NRs no noticeable effects
arise. Electron and hole remain in the center of the rod until the field splits them 
up towards opposite NR ends (not shown). Conversely, type-II NRs display an interesting
interplay between the electric field and Coulomb polarization effects, whose effect
on the charge distribution is summarized in figure~\ref{Fig6}.
A small electric field ($F=20\;kV/cm$) suffices to localize the hole in the CdTe 
shell near the negative electrode.  The electron localization is however strongly
dependent on the dielectric environment. In the absence of dielectric mismatch
($\varepsilon_{\mathrm{out}}=9.2$) it is centered, revealing a compensation between the
electric field and electron-hole interactions. For $\varepsilon_{\mathrm{out}}=2$,
Coulomb interaction dominates and the electron moves towards the hole (in spite
of the electric field), and the opposite occurs for $\varepsilon_{\mathrm{out}}=25$.
With increasing electric field ($F=100\;kV/cm$), the electron is forced to
move towards the positive electrode, but this is still difficult if the
environment is strongly insulating ($\varepsilon_{\mathrm{out}}=2$).
Once again, this behavior comes from the modulation of the exciton binding energy 
by the dielectric confinement.

\section{Conclusions}

We have shown that the dielectric confinement has significant effects in the
excitonic properites of semiconductor NRs. In type-I NRs, low dielectric 
constant environments blueshift the exciton photoluminescence peak by tens of meV, 
enhance electron-hole recombination rates and increase the electric field required 
to separate electrons from holes. 
The two latter effects are direct consequences of the enhanced correlation regime and 
exciton binding energy, while the former is a consequence of the exciton self-interaction with 
the induced polarization charges.

In type-II NRs, the same effects hold, but now greatly enhanced due to the
asymmetric charge distribution of electrons and holes, which reduces the
compensation between self-interaction and electron-hole Coulomb polarization.
In these systems, a strong dielectric mismatch may move the electron charge
density from the center of the core towards the heterostructure interface. 
This result has straightforward implications in the physical response of the NRs, 
and it shows that the dielectric confinement can be used -in addition to spatial confinement- 
to manipulate the shape and size of type-II excitons. 

To experimentally confirm the electronic density localization trends reported here, we
propose using wave function mapping techniques, such as near-field scanning optical microscopy~\cite{MatsudaPRL}.
Alternatively, the diamagnetic shift of NRs subject to transversal magnetic fields will discriminate
excitons with an electron localized in the center or near the shells of the NR.
We close by noting that the phenomena reported in this work are not exclusive of CdSe/CdTe NRs. 
They can be extended to rods made of different materials as long as the appropiate dielectric
confinement regime is attained.

\ack{Support from MCINN project CTQ2008-03344, UJI-Bancaixa project P1-1A2009-03,
a Generalitat Valenciana FPI grant (MR) and the Ramon y Cajal program (JIC)
is acknowledged.}

\section*{References}
\bibliography{NR}
\bibliographystyle{iopart-num}

\end{document}